\documentclass[epj]{svjour}

\input epsf

\newcommand{\be}{\begin{equation}}
\newcommand{\ee}{\end{equation}}
\newcommand{\bea}{\begin{eqnarray}}
\newcommand{\eea}{\end{eqnarray}}
\newcommand{\g}{\gamma_1}
\newcommand{\gb}{\gamma_{11}}
\newcommand{\wed}{(\pi/2)}
\newcommand{\fla}{(\pi)}

\newcommand{\go}{\gamma_1^{{\rm o}}}
\newcommand{\gs}{\gamma_1^{{\rm s}}}
\newcommand{\gm}{\gamma_1^{{\rm m}}}
\newcommand{\kl}{\kappa_l}
\newcommand{\kr}{\kappa_r}
\newcommand{\kv}{\kappa_v}
\newcommand{\kh}{\kappa_h}

\def\ffrac#1#2{\textstyle{#1\over#2}\displaystyle}

\begin{document}

\title{Two-dimensional polymer networks at a mixed boundary:\\
Surface and wedge exponents}
\authorrunning{M. T. Batchelor et al.}
\titlerunning{Polymers at a mixed boundary}
\author{M. T. Batchelor\inst{1}, D. Bennett-Wood\inst{2} \and 
A. L. Owczarek\inst{2} 
}                     
\institute{Department of Mathematics, School of Mathematical
               Sciences, The Australian National University,\\
                Canberra ACT 0200, Australia \and
     Department of Mathematics and Statistics, 
The University of Melbourne, Parkville, Victoria 3052, Australia}

\date{\today}

\abstract{
We provide general formulae for the configurational exponents of an
arbitrary polymer network connected to the surface of an arbitrary
wedge of the two-dimensional plane, where the surface is allowed to
assume a general mixture of boundary conditions on either side of the
wedge. We report on a comprehensive study of a linear chain by exact
enumeration, with various attachments of the walk's ends to the
surface, in wedges of angles $\pi/2$ and $\pi$, with general mixed
boundary conditions.
} 

\maketitle


\section{Introduction}\label{intro}

The configurational properties of long self-avoiding polymer chains in
the vicinity of a boundary has long been of interest \cite{BGMTW78}.
Recent progress has involved combining general results from scaling
and conformal invariance \cite{DL93,E93,C96} with results from exactly
solved lattice models.
The canonical model of polymers in a solvent is that of self-avoiding
walks (SAWs) on a lattice.  A wall can be introduced by
restricting the SAW to the upper half of the lattice, and
the interaction with the surface by an energy, $\varepsilon$,
associated with contacts between the polymer and the surface.  The
Boltzmann weight for a configuration of the polymer is given by
$\kappa^m = e^{ m \varepsilon /k_B T}$, where $T$ is the temperature of the
solvent and $m$ is the number of contacts with the surface.  At some
critical temperature, $T_a$, the polymer becomes adsorbed onto the
surface \cite{hamm-torrie-whitt-82}. For high temperatures, $(T>T_a)$,
the polymer is in a desorbed phase where it extends a large distance
into the solvent above the surface to which it is attached.  For low
temperatures, $(T<T_a)$, the polymer is in an adsorbed phase. It is
well known \cite{deGennes-79} that there is a correspondence between
SAWs and the $O(n)$ model in the limit $n \rightarrow 0$. The $O(n)$ model
has been considered with three different boundary conditions:
free boundary spins, where the bulk and surface couplings are the same;
fixed boundary spins; and critically enhanced surface
coupling \cite{burk-eisen-94}. In the terminology of surface critical
phenomena these three boundary conditions correspond to the `ordinary',
`extraordinary' and `special' transitions.  The critical adsorption
temperature, $T_a$, for SAWs corresponds to the `special' transition,
whilst the `ordinary'
transition corresponds to SAWs in the presence of an effectively
repulsive surface.

        Recently Batchelor and Yung \cite{BY95b} derived the critical
temperature and
configurational exponent from the Bethe Ansatz solution of the $O(n)$
loop model with mixed boundary conditions on the honeycomb lattice.
Here, `ordinary' (o) boundary conditions apply to one side of the
walk's origin, and `special' (s) boundary conditions apply to the
other. The general model with a flat surface and mixed boundary
conditions on the honeycomb lattice has been discussed by Bennett-Wood
and Owczarek \cite{BO96} who verified the critical temperature and
exponent values.

        Here we provide the general formulae for the
configurational exponents of an arbitrary polymer network connected
to the surface of an arbitrary wedge of the two-dimensional plane where
the surface is allowed to have general mixed boundary conditions.
We also report on an extensive numerical study of this situation.
This confirms the theory and brings together
consistent numerics for all previously studied cases. Our results
are given in Table~1.


\section{Surface exponents for arbitrary mixed topology}

\begin{figure}
\centerline{
\epsfxsize=2.5in
\epsfbox{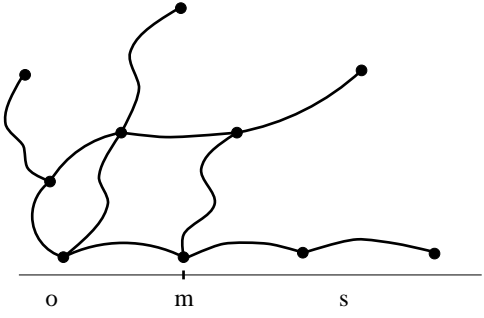}}
\caption{A network made of ${\cal N} = 11$ chains at the mixed o-s
boundary. There are $V=6$ bulk vertices, with
$n_1 = 3$, $n_3 = 2$, $n_4 = 1$ and $V'=4$ surface vertices,
with $n_3^{\rm o} = 1$,
$n_3^{\rm m} = 1$, $n_1^{\rm s} = 1$, $n_2^{\rm s} = 1$.}
\label{net}
\end{figure}

We consider the most general mixed network of $\cal N$ identical long
self-avoiding polymer chains of lengths $S$. Each chain ends in a
vertex.  The surface geometry is depicted in Fig.~\ref{net}. Vertices
in the vicinity of the boundary where the boundary conditions
change from o to s are denoted m.
The topological characteristics are the
numbers $n_L$ of $L$-leg vertices in the bulk and $n'_L$ of
$L$-leg vertices near the surface. In particular,
$n'_L$ is the total number of vertices of each type,
$n'_L = n_L^{\rm o} + n_L^{\rm m} + n_L^{\rm s}$.
In each case there can be $L \ge 1$ vertices. The total
number of bulk and surface vertices are given by
$V = \sum n_L$ and $V' = \sum n'_L$. The number of chains
can be written as
$ {\cal N} = \frac{1}{2} \sum ( n_L + n'_L) L $.

The number of configurations ${\cal Z}_G$ of the network is given by
${\cal Z}_G \sim \mu^{{\cal N} S} S^{\gamma_G -1}$ as $S \to \infty$.
Here $\mu$ is the connective constant for SAWs.
The general argument for the universal exponent $\gamma_G$ follows that
given in \cite{DS86} for the polydisperse partition function, with result
\begin{eqnarray}
\gamma_{G {\rm poly}} &=& \nu \left[ \phantom{\sum_L} \!\!\!\!\!
2 V + V' - 1   \right. \nonumber\\
 &-& \left. \sum_L
     (n_L x_L + n_L^{\rm o} x_L^{\rm o} + n_L^{\rm m} x_L^{\rm m} +
n_L^{\rm s} x_L^{\rm s} ) \right] .
\end{eqnarray}
We also make the assumption that the chains are monodisperse, with
$
\gamma_G = \gamma_{G {\rm poly}} - {\cal N} + 1.
$
Collecting the terms, with $\nu = \frac{3}{4}$ \cite{N82}, then gives
\begin{eqnarray}
\gamma_G = \ffrac{1}{4} &-& \sum_L \left(
      \ffrac{3}{4} x_L + \ffrac{1}{2} L - \ffrac{3}{2} \right) n_L
            \nonumber \\
        &-& \sum_L \left(
      \ffrac{3}{4} x_L^{\rm o} + \ffrac{1}{2} L - \ffrac{3}{4} \right)
            n_L^{\rm o}  \nonumber \\&-&  \sum_L \left(
      \ffrac{3}{4} x_L^{\rm m} + \ffrac{1}{2} L - \ffrac{3}{4} \right)
          n_L^{\rm m} \nonumber \\ &-& \sum_L \left(
    \ffrac{3}{4} x_L^{\rm s} + \ffrac{1}{2} L - \ffrac{3}{4} \right)
            n_L^{\rm s}  \, , \label{gi}
\end{eqnarray}
where the $x_L$ are geometric scaling dimensions. These have
all been derived for the bulk, ordinary, special and mixed
transitions \cite{BB89,BS93,BY95a,BY95b,YB95b} from the exactly
solved $O(n)$ model on the honeycomb lattice \cite{B86,BS93,YB95a}.
The dimensions $x_L$ and $x_L^{\rm o}$ had been obtained earlier
by conformal invariance and Coulomb gas methods \cite{N84,Sa,D,C84a,DS86}. 
In particular, at $n=0$ 
\begin{eqnarray}
x_L &=& \ffrac{3}{16} L^2 - \ffrac{1}{12}, \nonumber \\ 
x_L^{\rm o} &=& \ffrac{3}{8} L^2 + \ffrac{1}{4} L, \nonumber \\ 
x_L^{\rm s} &=& \ffrac{3}{8} (L+1)^2 - \ffrac{3}{2}(L+1) +
\ffrac{35}{24}, \nonumber \\ 
x_L^{\rm m} &=& \ffrac{3}{8} L^2 - \ffrac{1}{4} L .
\end{eqnarray}
Inserting these results into (\ref{gi}) gives
\begin{eqnarray}
\gamma_G = \ffrac{1}{4} &+& \ffrac{1}{64} \sum_L  n_L (2-L) (9L+50)
\nonumber \\
                       &-& \ffrac{1}{32} \sum_L n_L^{\rm o}
       (9L^2 + 22 L - 24) \nonumber \\
   &-& \ffrac{1}{32} \sum_L n_L^{\rm m} (9L^2 + 10 L - 24)
\nonumber \\
   &-& \ffrac{1}{32} \sum_L n_L^{\rm s} (9L^2 -2 L - 16) \, .
   \label{g}
\end{eqnarray}
The exponents for a pure `ordinary' surface \cite{DS86} are recovered
with $n_L^{\rm m} = n_L^{\rm s} = 0$.
For mixed boundaries there is only one $L$-leg vertex emanating
from the origin, thus $n_L^{\rm m} = 1$.

\section{Wedge exponents}

\begin{figure}
\centerline{
\epsfxsize=3.5in
\epsfbox{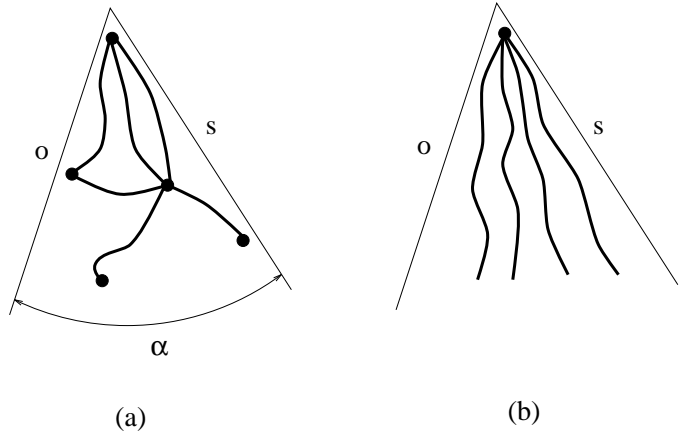}}
\caption{(a) A mixed o-s network with a vertex in a wedge of angle $\alpha$;
(b) a star polymer.}
\label{wedge}
\end{figure}

The network can be tied in a wedge of angle $\alpha$ by an $\hat L$-leg
vertex as in Fig.~\ref{wedge}. Obtaining the wedge network exponents
$\gamma_G(\alpha)$
involves a conformal map of the wedge to the half-plane \cite{C84a,DS86}.
The final result
\be
\gamma_G(\alpha) = \gamma_G(\pi) - \nu \left( \ffrac{\pi}{\alpha} -1
\right) x'_{\hat L} \label{w}
\ee
is as given in \cite{DS86}, where now $\gamma_G(\pi)$ is the half-plane
exponent (\ref{g}). The contribution $x'_{\hat L}$ from the $\hat L$-leg
tie depends on the particular surfaces under consideration,
with $x'_{\hat L} = x^{\rm m}_{\hat L}$ for the mixed boundary.

It follows from (\ref{g}) and (\ref{w}) that an $L$-leg star polymer
confined to a wedge with o-o, s-s or o-s surfaces has exponents
\bea
\gamma_L^{\rm o}(\alpha) &=& 1 + \frac{27 L}{64} -
\frac{3 \pi L (3L+2)}{32 \alpha} \, , \label{ds} \\
\gamma_L^{\rm s}(\alpha) &=& 1 + \frac{27 L}{64} -
\frac{ 9 \pi L (L-2) + 8 \pi}{32 \alpha} \, , \\
\gamma_L^{\rm m}(\alpha) &=& 1 + \frac{27 L}{64} -
\frac{3 \pi L (3L-2)}{32 \alpha} \, .
\eea
The o-o result (\ref{ds}) is that obtained in \cite{DS86}.
As particular examples relevant to our numerical tests, consider
a single SAW emanating from a $90^\circ$ corner.
In this case the above formulae reduce to
$\gamma_1^{\rm o}(\ffrac{\pi}{2}) = \ffrac{31}{64}$,
$\gamma_1^{\rm s}(\ffrac{\pi}{2}) = \ffrac{95}{64}$,
$\gamma_1^{\rm m}(\ffrac{\pi}{2}) = \ffrac{79}{64}$.
The exponents differ if the walk terminates on either boundary.
In that case (\ref{g}) and (\ref{w}) give
$\gamma_{11}^{\rm o}(\ffrac{\pi}{2}) = -\ffrac{21}{32}$ for the o-o corner
and $\gamma_{11}^{\rm s}(\ffrac{\pi}{2}) = \ffrac{27}{32}$ for the s-s corner.
For the o-s corner the walk can terminate on either the o side,
with $\gamma_{11}^{\rm mo}(\ffrac{\pi}{2}) = \ffrac{3}{32}$,
or on the s side, with $\gamma_{11}^{\rm ms}(\ffrac{\pi}{2}) = \ffrac{19}{32}$.


\section{Adsorption temperatures on the honeycomb lattice}

There are two regular types of boundary of a honeycomb lattice. For
the horizontal boundary of Fig.~\ref{honeywalkwedge} the critical
adsorption temperature is known to be given by \cite{BY95a}
\be
\exp \left( \frac{\varepsilon}{k T_a} \right) =
{1 + \sqrt 2} = 2.414 \ldots \label{ctempone}
\ee
This result follows from the boundary vertex weights of the
corresponding exactly solvable $O(n)$ loop model \cite{YB95a}.

The adsorption temperature for the vertical boundary of
Fig.~\ref{honeywalkwedge} can be determined in a similar
way. Specifically, the horizontal boundary in Fig.~\ref{honeywalkwedge}
follows on taking the value $u=\lambda$ in the vertex weights of the
more general solvable loop model on the square lattice \cite{BY95a}.
The vertical boundary in Fig.~\ref{honeywalkwedge} follows on taking the value
$u=2\lambda$. As a result the critical fugacity of a step along the
boundary is given by
$y^* = 1/\sqrt{t_b t_s}$
where
$t_b = 2 \cos \ffrac{\pi}{8} = (2 + \sqrt 2)^{1/2}$
and $t_s^2 = {\cos \ffrac{5\pi}{16}}/{\cos\ffrac{\pi}{16}} = t_b^2 - t_b -1$.
The critical adsorption temperature is thus given by
\be
\exp \left( \frac{\varepsilon}{k T_a} \right) =
{{\sqrt{{\frac{2 + {\sqrt{2}}}{1 + {\sqrt{2}} -
{\sqrt{2 + {\sqrt{2}}}}}}}}} = 2.455 \ldots \label{ctemptwo}
\ee


\section{Results from exact enumeration}

The general theory presented above generalises that already given for
polymer networks in wedges of arbitrary angles \cite{DS86} to the case
of mixed boundary conditions. To test the exponent predictions in this
case we have enumerated SAWs on the honeycomb lattice with various
ends attached to a surface and confined in wedges of two different
angles ($\pi/2$ and $\pi$). This also allows us to verify the critical
boundary temperature (\ref{ctemptwo}). The critical temperature
(\ref{ctempone}) and the exponent $\gamma_1^m(\pi)$ have been verified
previously \cite{BO96}. (A comprehensive account of this study can be
found in \cite{thesis}.)  Previous numerical work other than
\cite{BO96} has focused on the square and triangular
lattices. However, our numerical task is made considerably easier since
the exact prediction for the connective constant for SAWs
on the honeycomb lattice, $\mu =\sqrt{2 + \sqrt 2}$, allows the
biasing of exponent estimates.

\begin{figure}
\centerline{
\epsfxsize=2.5in
\epsfbox{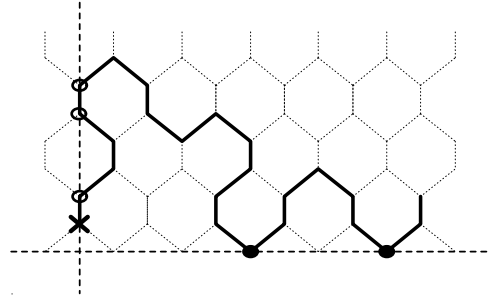}}
\caption{A SAW (origin denoted by a cross) on the
honeycomb lattice  attached to the
vertex of a $90^\circ$ wedge with contacts identified with the vertical
 boundary (open circles) and with the horizontal boundary (closed circles).
 A fugacity $\kappa_v$ is associated with open circle
contacts while a fugacity $\kappa_v$ is associated with
closed circle contacts.  (The walk is not permitted to go beyond
the indicated dotted lines.)}
\label{honeywalkwedge}
\end{figure}

In this letter we consider two particular situations: SAWs restricted
to the upper half plane of the honeycomb lattice with fugacities $\kl$
and $\kr$ associated with contacts between the walk and either side of
the surface as shown in Fig. 1 of \cite{BO96}, and more importantly
SAWs restricted to the positive quadrant of the honeycomb lattice,
where the fugacities $\kv$ and $\kh$ are associated with contacts
between the walk and the vertical and horizontal surfaces, respectively,
as shown in Fig.~\ref{honeywalkwedge}.
The partition function for walks of length $n$ with one end attached
to a surface, with different energies for sites interacting with
either side of a $\pi/2$ wedge (see Fig.~\ref{honeywalkwedge}), is
given by
\be
Z_n^1(\kv,\kh,\pi/2) = \sum_{m_v,m_h} c^1_n(m_v,m_h,\pi/2)\kv^{m_v}
\kh^{m_h} \; ,
\ee
where the sum is over all allowed values of the number of contacts
$m_v$ with the vertical boundary, and the number of contacts $m_h$
with the horizontal boundary. The coefficients $c^1_n(m_v,m_h,\pi/2)$
are the number of configurations of length $n$ with $m_v$ and $m_h$
vertical and horizontal interactions respectively. The partition
functions for walks with the both ends attached,
$Z_n^{11}(\kv,\kh,\pi/2)$, simply replaces $c^1_n(m_v,m_h,\pi/2)$ with
$c^{11}_n(m_v,m_h,\pi/2)$ for configurations attached at both ends.
Also, in the cases, as described in \cite{BO96}, associated with a
flat surface (wedge angle $\pi$), the partition functions,
$Z_n^1(\kl,\kr,\pi)$ and $Z_n^{11}(\kl,\kr,\pi)$, are defined with the
obvious modifications. Tables of the various coefficients
$c^1_n(m_v,m_h,\pi/2)$, etc.\ can be found in
\cite{thesis}\footnote{or via email to aleks@ms.unimelb.edu.au}. Cases
where walks traverse from one side of the wedge to the other have also
been considered but numerical difficulties hampered exponent
calculation and hence we do not present those results here
\cite{thesis}. The method of enumeration of the coefficients was via a
backtracking algorithm, which was implemented (on a small
supercomputer) in a similar way to that described in \cite{BO96}.

By setting the interaction parameters to the particular values, e.g.\
those implied from the critical temperature values given in the
previous section, estimates of the various exponents were obtained by
analysis of the singularity structure of generating functions of the
resulting partition functions. The method of analysis was based
primarily on biased differential approximants as explained in
\cite{BO96}.

For completeness we give estimates of all the exponents
$\gamma_1(\pi)$ and $\gamma_{11}(\pi)$, and $\gamma_1(\pi/2)$ and
$\gamma_{11}(\pi/2)$ for each boundary condition. In most cases the
numerical accuracy equals or surpasses previous estimates. It should
be noted though that errors quoted are \emph{not} rigorous bounds and
represent the spread of the approximants' exponent values (at the
critical point). The difference in the accuracy (some values are more
accurate, e.g.\ $\go\fla$, than the errors --- which are
conservative --- suggest) can be argued to be due to the amount of
shift required in biasing the approximants, which is itself an
indication of the relative convergence of the series to the asymptotic
forms expected. A full discussion of how the errors and final
estimates were obtained can be found in \cite{thesis}, following the
general lines given in \cite{BO96}. Our estimates for all exponents
are in good agreement with the predicted values and are given in
Table~1. The verification of the special exponents
involved the implicit verification of the vertical adsorption
temperature, eqn.~(\ref{ctemptwo}).

\begin{table}
\caption{
Our estimates for the entropic exponent $\gamma$ for SAWs
attached to a flat surface or $\pi/2$ wedge.}
\begin{tabular}{ll}
\hline\noalign{\smallskip}
numerical & predicted  \\
\noalign{\smallskip}\hline\noalign{\smallskip}
$\go\fla = 0.9531(5)$&$(\frac{61}{64}=0.953125)$\\
\noalign{\smallskip}
$\gb^{\rm o}\fla = -0.186(2)$&$(-\frac{3}{16}=-0.1875)$\\
\noalign{\smallskip}
$\g^{{\rm o}}\wed=0.4843(9)$&$(\frac{31}{64}=0.484375)$\\
\noalign{\smallskip}
$\gb^{{\rm o}}\wed=-0.655(3)$&$(-\frac{21}{32}=-0.65625)$\\
\noalign{\smallskip}\hline\noalign{\smallskip}
$\gs\fla=1.451(2)$&$(\frac{93}{64}=1.453125)$\\
\noalign{\smallskip}
$\gb^{\rm s}\fla= 0.813(4)$&$(\frac{13}{16}=0.8125)$\\
\noalign{\smallskip}
$\g^{{\rm s}}\wed=1.482(8)$&$(\frac{95}{64}=1.484375)$\\
\noalign{\smallskip}
$\gb^{{\rm s}}\wed=0.85(1)$&$(\frac{27}{32}=0.84375)$\\
\noalign{\smallskip}\hline\noalign{\smallskip}
$\gm\fla=1.3279(5)$&$(\frac{85}{64}=1.328125)$\\
\noalign{\smallskip}
$\gb^{\rm mo}\fla=  0.183(6)$&$(\frac{3}{16}=0.1875)$\\
\noalign{\smallskip}
$\gb^{\rm ms}\fla= 0.689(9)$&$(\frac{11}{16}=0.6875)$\\
\noalign{\smallskip}
$\g^{\rm m}\wed=1.233(6)$&$(\frac{79}{64}=1.234375)$\\
\noalign{\smallskip}
$\gb^{\rm mo}\wed= 0.09(1)$&$(\frac{3}{32}=0.09375)$\\
\noalign{\smallskip}
$\gb^{\rm ms}\wed= 0.596(7)$&$(\frac{19}{32} = 0.59375)$\\
\noalign{\smallskip}\hline
\end{tabular}
\end{table}


\section{Conclusion}

We present the general results for the entropic exponents of a polymer
network in two dimensions attached to the surface in a general wedge
topology. We have verified that the theoretical formulae, coming from
a combination of scaling and conformal invariance considerations and
exact results, are correct by extensively analysing exact enumeration
data from SAWs on the honeycomb lattice. Where numerical evidence has
been precise confirmation of the theory has been good. However, several
questions remain. One is the numerical confirmation of exponents in
this general setting for more complicated examples. Another
associated question is the validity of one of the assumptions of the
theory concerning polydisperse verses monodisperse cases \cite{last}. 
Such studies are outside the range of current exact enumeration  
and probably require careful Monte Carlo work.

\section*{Acknowledgments}

It is a pleasure to thank John Cardy for explaining the
general scaling and conformal invariance arguments to us.
Two of us (MTB and ALO) have been supported by the Australian
Research Council.



\begin{thebibliography}{99}

\bibitem{BGMTW78} M.~N. Barber, A.~J. Guttmann, K.~M. Middlemiss,
G.~M. Torrie and S.~G. Whittington, J. Phys. A {\bf 11}, (1978) 1833.

\bibitem{DL93} K. De'Bell and T. Lookman, Rev. Mod. Phys.
               {\bf 65}, (1993) 87.

\bibitem{E93} E. Eisenriegler, Polymers Near Surfaces
              (World Scientific, Singapore, 1993).

\bibitem{C96} J. Cardy, Scaling and Renormalization in Statistical Physics
              (CUP, Cambridge, 1996).

\bibitem{hamm-torrie-whitt-82}
J.~M. Hammersley, G.~M. Torrie and S.~G. Whittington,
J. Phys. A {\bf 15}, (1982) 539.

\bibitem{deGennes-79} P.~G. de~Gennes, Scaling concepts in polymer physics
(Cornell University, 1979).

\bibitem{burk-eisen-94}
T.~W. Burkhardt and E.~Eisenriegler, Nucl. Phys. B {\bf 424}, (1994) 487.

\bibitem{BY95b} M.~T. Batchelor and C.~M. Yung, J. Phys. A
                {\bf 28}, (1995) L421.

\bibitem{BO96} D. Bennett-Wood and A.~L. Owczarek,
               J. Phys. A {\bf 29}, (1996) 4755.

\bibitem{DS86} B. Duplantier and H. Saleur, Phys. Rev. Lett.
               {\bf 57}, (1986) 3179; B. Duplantier, J. Stat. Phys. 
{\bf 54}, (1989) 581.

\bibitem{N82} B. Nienhuis, Phys. Rev. Lett.
              {\bf 49}, (1982) 1062.

\bibitem{BB89} M.~T. Batchelor and H.~W.~J. Bl\"ote,
               Phys. Rev. B {\bf 39}, (1989) 2391.

\bibitem{BS93} M.~T. Batchelor and J. Suzuki,
               J. Phys. A {\bf 26}, (1993) L729.

\bibitem{BY95a} M.~T. Batchelor and C.~M. Yung,
                Phys. Rev. Lett. {\bf 74}, (1995) 2026.

\bibitem{YB95b} C.~M. Yung and M.~T. Batchelor, Nucl. Phys. B
                {\bf 453}, (1995) 552.

\bibitem{B86} R.~J. Baxter, J. Phys. A {\bf 19}, (1986) 2821.

\bibitem{YB95a} C.~M. Yung and M.~T. Batchelor,
                Nucl. Phys. B {\bf 435}, (1995) 430.

\bibitem{N84} B. Nienhuis, J. Stat. Phys. {\bf 34}, (1984) 731.

\bibitem{Sa} H. Saleur, J. Phys. A {\bf 20}, (1987) 455; {\bf 19}, 
(1986) L807.

\bibitem{D} B. Duplantier, Phys. Rev. Lett. {\bf 57}, (1986) 941.

\bibitem{C84a} J.~L. Cardy, Nucl. Phys. B {\bf 240}, (1984) 514.

\bibitem{thesis} D. Bennett-Wood, Ph.\ D.\ Thesis, University of Melbourne (1998).

\bibitem{last} At some length this is discussed in L. Sch\"afer, C. von Ferber,
U. Lehr and B. Duplantier, Nucl. Phys. B {\bf 374}, (1992) 473. 

\end{thebibliography}
\end{document}